\documentclass{./LLNCS/llncs}

\usepackage[english]{babel}
\usepackage[T1]{fontenc}

\usepackage{pstricks}
\usepackage{fancybox}
\usepackage[dvips]{graphics}
\usepackage{./zedbis}
\usepackage{xspace}

\usepackage{boxedminipage}
\usepackage{ae}

\def\kmelia{\textsf{Kmelia}\xspace}

\title{Can Component/Service-based Systems be Proved Correct?}
\titlerunning{Software Services Verification}

\author{J. Christian Attiogb{\'e} 
}
\institute{LINA  - UMR CNRS 6241 - University of Nantes - France\\
\texttt{Christian.Attiogbe@univ-nantes.fr} \\[0.7ex]
(Technial Report version)\\
}

\begin{document}

\maketitle

\begin{abstract}
Component-oriented and  service-oriented approaches have gained a strong
 enthusiasm in industries and academia with a particular interest for
 service-oriented approaches. 
The service concept and its use in 
 web-based application development have a huge impact on reuse practices. 
Accordingly a considerable part of software architectures is influenced; these architectures are moving towards
 service-oriented architectures. 
Therefore applications
 (re)use services that are available elsewhere and many applications interact, without knowing each other, using
 services available via service servers and through their published interfaces and
 functionalities.

Industries propose, through various consortium,  languages, technologies and
 standards. More academical works are also undertaken concerning semantics
 and formalisation of components and service-based systems.

We consider here both streams of works in order to raise research
 concerns that will help in building quality software.

Are there new challenging problems with respect to
 service-based software
 construction.
\end{abstract}

\section{Introduction}

Component-oriented and  service-oriented approaches have gained a strong enthusiasm in industries and academia with a particular interest for service-oriented ones. 

A component is a software entity with given functionalities, made
 available by a provider, and used to build other application within
 which it is integrated.
 The concept of service as \textit{a specific software entity which is
 delivered upon request via a network} is not a new idea; think about 
telecommunication services, IP network services, etc.

A service is a software entity that is designed for a given
 functionality, is made available through a network via a provider, and is linked and used
 on demand. An example is a calendar
 service used within a flight booking application in order to help the
 user in defining its flight departing  and returning dates. The
 calendar service is a service  integrated in the booking application;
 it may be replaced by another one or may even be
 disconnected, in which case the user may for example enter  manually her/his dates.    

The service concept and its use in 
 web-based application development have a huge impact on reuse practices. 
Accordingly a considerable part of software architectures is influenced; these architectures are moving towards
 service-oriented architectures (SOA). 
Applications  (re)use services that are
available elsewhere and several  applications interact, without knowing each other, using
the services available via service servers and their published interfaces and
 functionalities.

 The application designers are confident in the used services and their
 (unknown) providers; they are also confident in the used communication protocols,  
 even if the services may have misleading semantics. 
 Important issues at this stage are the description, the storage, the
 management and the retrieval of appropriate services from large
 \textit{certified} libraries. Standardisation effort is therefore
 unavoidable to ensure certification.

Software service descriptions take various forms; services are described
 either inside a software
 component which role is to provide a computation, or inside a software
 component which provides several other services as functionalities; 
 services may be described using  software components.

When an external service is considered within a more
 general software
 application, one should think about the meaning of the interaction
 between the application and the external service and, consequently, the meaning of the application
 with or without the links with these external services. It is worth
 mentioning a few
 key features: service description, service discovery, service interaction,
 service-oriented applications and their architectures.
In  \cite{DBLP:conf/ictac/Fiadeiro04} Fiadeiro advocates for the emerging of a
new discipline concerning software services.
Papazoglou and van den Heuvel propose a general panel in \cite{PapazoglouVLDB2007} of the
technology around service-oriented architecture. An accompanying work
can be found in \cite{PapazoglouCIS2008}.

Are there new challenging problems with respect to
 service-based software
 construction, to service construction, to software verification?
 Besides, what  are the links and the advances compared to distributed systems? 
Do the technical proposals meet the industry expectations?
what are the current trends and the perspectives of service-oriented approaches?

Considering the ubiquitous aspect of large scale distributed applications, the way software services are impacting software
 construction or  the way they are being
 used or researched requires special care.
The standardisation of service description languages constitutes an
important step but  is not yet sufficient. The meaning of services, their
constraints and  their properties are unavoidable to guarantee the
well-functioning of global distributed applications that use them.

The main message stressed here, is that there  is a real
need to think about services and components as software building units
that should last several years or decades and that require for this
purpose technical specifications and the guarantee of
well-functioning. Indeed today Information Systems and Applications, due
to the Internet, are inter-penetrated in such a way that failures or
defects of some components or services may drastically impact  users and
their activities in the future.

Then, a specific emphasis should be put on correctness properties of services
and on service-based systems in order to ensure their quality.
An important research issue is to reason on the correctness of
software applications that will dynamically use or embed existing
services provided by a third party.  For example, additionally to  the formal specification of its
functionalities, a service  may embed its specific properties and the
certificate or the proof that  guarantees these properties.

From the methodological point of view, any given "$\alpha$-oriented"
development method requires a clear  description of what $\alpha$
is or supposed to be; therefore helping to  differentiate an $\alpha$-oriented approach from a $\beta$-oriented one.

An $\alpha$-oriented approach should make it explicit some  
concepts, laws, rules,  guidance or method to apply it distinctly from $\beta$-oriented approach even if an $\alpha$-oriented approach may be based on
 $\beta$-oriented ones with respect to methodological
 purpose. From this point of view, it should
 be clear how to situate service-oriented  development relatively to
 distributed-system development. 
 
In the object-oriented approach for example, one considers the
encapsulation of data   and the methods that manipulate the data as
elementary units (classes) which are related  through inheritance, extension and  client-ship.
 
From this point of view, service-oriented approaches require a clear
positioning that is currently disseminated through various existing works.
However a systematic view that will favour  the emergence of mature
seamless  development process and integrated tools and also academic courses is recommendable.

According to a given business application  a software functionality
is a specific need which occurs time being, then an access and a link to the
corresponding service is achieved on demand. The service should be
first identified and the located before being used.

A current widely admitted practice of service-based approaches is that
services are available somewhere in the Internet; they are accessed on
request and  a given  application may perform, during its execution
access to the available  services. Therefore it is worth underlying the
issue on service correctness and interoperability. All these are
challenging concerns. The standardisation of service descriptions including
 their specifications  (interfaces, behaviour and semantics), their
 properties and their  quality is the only way to overcome these challenges.

The remaining of this article is structured around four sections.
Section \ref{section:CompoServi} is devoted to the introduction of the
well-admitted notion of components and services.
In Section \ref{section:analysis} we deal with the analysis of
component or
services based systems. Section \ref{section:researchIssue} is devoted to
research issues and related directions. Finally in Section
\ref{section:conclu} we give some concluding remarks.

\section{Components and Services in Software Construction}
\label{section:CompoServi}
\vspace{-0.3cm}
\subsection{Reuse of Software}
The reuse principle has always been present as a fundamental support for
software construction. Functions, Modules and Classes have been the
concrete entities that enforce reuse. In the same stream, Components and
Services follow on, and are just the adaptation of the reuse practices to the
environment of software construction. Indeed one may observe the
evolution of the reuse
entities from the point of view of centralised or decentralised
environment. Functions, Modules and Classes are well-adapted to
centralised environment. Libraries of these entities are available \textit{in situ} for
the software developers. With the maturity of network technologies and
mainly the Internet protocols (via the TCP/IP facilities) the
abstraction level of reuse entities changes: Component-based
development appears in this context in the 90's.
Clearly as the software technological environment changes, abstraction
level of reuse entities changes.
Services are now, the most widely highlighted reuse entities.

\vspace{-0.3cm}
\subsection{Comparing Components and Services}

Components are viewed as the abstraction of a part or a subsystem of an
application. The notion of component is widely used in many disciplines
and industries. For example the Hardware in a computer is
made of several linked components (the CPU, input-output devices,
storage units, etc). 
A computer is made of several components that interact at different
levels (within the Hardware, the operating system and in the applications). 

Therefore a component is an independent building block that can be
independently integrated in one application or within several
applications. In a component-based approach  the functionalities of a
software are considered completely independent and designed as 
separate entities (the components) that can be composed afterthought in
building a applications. The components can be designed and offered by a
tier.

Services are software building blocks used in building distributed
applications. They are also used to integrate existing functionalities into
new applications, or used to make different distributed
applications communicate.
A service is provided  by a service provider and required by a consumer
or a service client; it can be dynamically located and invoked.
Therefore the network of computers, more specifically the Internet, is 
the support and the vehicle of services and consequently one of the main
characteristics of service-based systems.

\subsubsection{Web-Services}
Web-services are a specific implementation of services; they are now
very popular through the industrial offers. They use XML and the
Internet TCP/IP popular protocol.

According to the W3C\footnote{The World Wide Web Consortium}, \textit{A
(web)service is a software system designed to support interoperable
machine to machine interaction over a network.}

\medskip
Components and Services share many commonalities; they are both
independent entities used as building blocks in the construction of
software applications. They provide interfaces to enable usage and interactions.
However, a component is a packaging unit of one or
several services; therefore it provides one or several services.  A
service can be built using components but a service is not necessarily built from
components.
Services are clearly located at the functionalities abstraction level.
Services correspond to offered functionalities.

Components are known and assembled in order to enable the usage of
services. Components are integrated and compiled within new applications that use
them. They become a specific part of the new application.
Services are not compiled; they are dynamically linked at runtime within
the applications that use them.

Sharing aspect also particularise components and services.
A priori a component is not shared between  applications as the
component is integrated within the application. One service may
be shared by several clients. However there are variations on components
and services that do not fit this distinction, components or their
services may also be shared with specifically defined policies.   

From the service-based approach, interactions are not statically
predefined. The services  do not know their caller; service clients and 
the called services are bound upon request where clients behave like if
services are always available.

\vspace{-0.3cm}
\subsection{Service-Based Systems}

A service-based system is one that (re)uses services available elsewhere
in the Internet to achieve its own functionalities. Note that a
service-based system may be partially or totally architectured with
components. The components of the system request some functionalities
that are described as services outside the system. According to a user,
the system is completely independent and autonomous; the user does not
see the interactions on the background network.

Service-based systems are distributed systems: they are systems made of
collection of components (computers, software applications, devices,
etc); they are perceived by a user as a
single system. Their sub-systems are linked by a network equipped with
communication protocols, and  cooperate to share resources and to
perform their assigned tasks.
From the interaction point of view, services are facilities used by
software applications to make communicate various sub-systems connected via the Internet in order to achieve given tasks.

Compared with classical distributed systems, service-based systems have
the main features but are
rather unpredictable for that their parts may be unknown at a given
time. Indeed services are volatile distributed entities; they may be searched, dynamically linked with the rest
of the system environment, and unlinked at another moment.\\

\textit{Web-Services based applications} are service-based systems that use the Internet and its
protocols as the distribution platform. They are the widely used
 specialisation of service-based systems.

\vspace{-0.3cm}
\subsection{Event-based Style of Interaction}
An  event-based style of communication is well-adapted to the
construction of service-based systems.
In an event-based
style of interaction, the components of a distributed  application
communicate via events which are generated by ones components and received by others.   
This kind of interaction facilitates the decoupling of components. Indeed an
event notification system or an event management middle-ware handles the
interaction between the components of the application.
The components which are source or target of events are not specific
components. Therefore an event-based style favours the integration of
heterogeneous components and services in global applications.

\vspace{-0.3cm}
\subsection{Towards Standards for Service-Oriented Technology}

\subsubsection{Languages} 
\subsubsection{Web-Services Description Language (WSDL)}
The Web-Services Description Language  is one of today  \textit{de facto}
standard languages to define any kind of services. It is a markup language
based on the XML. It is imposed by the industry.

\subsubsection{Business Process Execution Language (BPEL)} The
(Web-Service) Business Process Execution Language\footnote{www-128.ibm.com/developerworks/library/specification/ws-bpel/}  is an
\textit{orchestration} language for web-services; it is used to describe
the interactions between web-services. BPEL can be simply viewed as a
language to compose services at abstract or execution levels.

\subsubsection{Protocols}
\subsubsection{Simple Object Access Protocol (SOAP)}
The Simple Object Access Protocol is an XML-based protocol to exchange  information (or messages) in
a distributed environment. It can  be used in combination with a variety of other
(Internet) protocols such as the HTTP protocol. It is also a protocol
from the W3C consortium.
Technically, the SOAP consists of three layers: 
\begin{itemize}
\item an external layer that defines a
framework for describing  \textit{message contents} and how to process them, 
\item a layer dedicated to  \textit{data encoding}; it describes a set of rules for expressing instances of application data, and 
\item a layer dedicated to the  \textit{representation convention} for remote procedure calls and responses. 
\end{itemize}

WSDL is used to describe services based on the SOAP.\\

\subsubsection{Universal Description, Discovery and Integration (UDDI)}
The Universal Description, Discovery and Integration (UDDI) is a protocol
over the HTTP protocol used to describe, locate and discover services.\\

Web-Services conform to these standards (WSDL, SOAP, UDDI).

\vspace{-0.3cm}
\subsection{Service-Oriented Approach and Architecture}

An architecture involving web-services uses three entities and four
relationships between them.
The involved entities  are:
\begin{itemize}
 \item a service (developer/provider),
 \item a service registry or server (service library),
 \item a client.
\end{itemize}

\noindent
The relationship are:
\begin{itemize} 
\item Service Registration, a service is registered on a server (service
 registry) ; the XML and HTTP protocol are used for that purpose;
 \item Service Discovery: a client looks for a service by consulting a
 service registry which in turn sends back the located service to the
       client; the Universal Description, Discovery Integration (UDDI)
       protocol is used for this purpose.
 \item Service Binding. A client is (dynamically) bound to the discovered
 service. Here, the Web-Service Description Language (WSDL) is used.
 \item Service Invocation. A client invokes a service to which it is
 previously bounded. The offered service  interacts with its client using the
 SOAP protocol. 
\end{itemize}

\medskip

\textbf{Service Oriented Architecture (SOA)}, originated from SUN company in the
90's. It is an architectural style to build applications that use the
services available on the Internet. The principle is that the services
are loosely coupled since they are not originally known but discovered
and bind upon demand; by the way, the components of an application are
also loosely coupled since the services to link them are used upon
demand. The SOA promotes the find, bind, use (services) method. It
favours the integration of heterogeneous distributed systems.
More details on SOA can be found in \cite{ErlSOA2005,PulierSOA2006}.

\section{Formal Analysis of Component/Service-Based Systems}
\label{section:analysis}
In the previous section we have considered the technologies of services
as imposed by industries.
Note that, considering problems to be solved, industrial time and
delays are different from academic ones. Industrial constraints are
also different; they often have to produce solutions with restricted
delays.

In this section, we consider the academic point of view with respect of
software development features, methods and constraints.
From this point of view, some stated properties  need to be verified for
software entities and applications. Verification is still a hard
activity that consumes a lot of time and resources (and expertise on formal
methods and tools).

Ongoing research efforts in the context of service-oriented applications
is for example the Sensoria FET European
project\cite{FiadeiroWS-FM2006}.
Related works and results on formalisation and semantics of services,
reasoning on web-services, and their coordination can be found in \cite{AbreuCoord2008,FiadeiroWS-FM2006,AbreuCoord2008,AbreuFMNet2007,FiaderoADT2007,BocchiWS-FM2007}.

\vspace{-0.3cm}
\subsection{Properties to be verified}
First of all software service construction requires formal
specifications in order to permit formal analysis. Expressive
specification languages with formal semantics and tools are needed.

The range of properties to be verified is very wide. To cite a few:
\begin{itemize}
\item availability of service (reliability of servers),
 \item functional properties of services,
\item reachability properties related to a given service or a functionality,
\item side-effect functional
       properties, for example a called service does not take the control forever, 
 \item correct interaction between a service an its clients;
    for example, the client is not blocked due to bad interaction with a
    requested service,
\item correct interaction between the services of an application,

 \item heterogeneity and interoperability questions; services may come
       from 
       various sources and have different semantic models;  how do they interact,
 \item correctness of message exchange protocols,
\item preservation of time-constraints
 \item $\cdots$
\end{itemize}

\medskip
There are a real need for component/service models to face specification
 and verification aspects.
\vspace{-0.3cm}
\subsection{The Kmelia Model Proposal}
We are experimenting on an abstract and formal component model named
\kmelia. It is a multi-service component model formally defined
\cite{coloss:sc06,coloss:sc08}. In the \kmelia  model a component has an
interface made of provided services and required services. Services are
used as  composition units. The provided services have an
interface made not only with usage information related to the signatures
but also with pre and post conditions in order to ensure at least formal
analysis. One strong idea of the model is that the development of
component and their services may follow a formal construction (from
abstract model to concrete code) in order
to enable property verification at different layers.  
\subsubsection{Service Specification}
\label{subsection:service}
A \kmelia service has the following shape:

\begin{center}
\begin{boxedminipage}{4cm}
\vspace{-0.3cm}
\begin{tabbing}
\=\hspace{0.7cm}\=\hspace{0.7cm}\=\hspace{1cm}\\
\>\textbf{{\sc service}} <serviceName(parameters)>\\
\>\> \textbf{{\sc interface}}\\
\>\>\> <dependencies with other (sub)services> \\
\>\> \textbf{{\sc properties}} <specific property references> \\
\>\> \textbf{{\sc pre}}  <a predicate> \\
\>\> \textbf{{\sc post}} <a predicate> \\
\>\> \textbf{{\sc behaviour}} \\
\>\>\> <An extended LTS, with initial and final states>\\
\>\textbf{{\sc end}}
\end{tabbing}
\end{boxedminipage}
\end{center}

\medskip
A \textit{service} $s$ of a component $C$ is defined with an \textit{interface} $I_s$ and a (dynamic) \textit{behaviour} ${\mathcal B}_s$:
$\langle I_s, {\mathcal B}_s\rangle$.
Usually a required service does not have the same level of detail as a provided service since a part of these details is already in the (provided) service that calls it.

The interface ${I}_s$ of a service $s$ is defined by a 5-tuple $\langle \sigma,\ P,\ Q,\ V_s, S_s \rangle$ where 
$\sigma$ is the service signature (name, arguments, result),

$P$ is a \textbf{precondition}, 
$Q$ is a \textbf{postcondition}, 
$V_s$ is a set of local declarations and 
the \textit{service dependency} $S_s$ is a 4-tuple $S_s = \langle
sub_s,\ cal_s,\ req_s,\ int_s\rangle$ of disjoint sets 
where $sub_s$ (resp. $cal_s$, $req_s$, $int_s$) contains the provided services names (resp. the services required from the caller, the services required from any component, the internal 
services) in the $s$ scope.
Using a required service $r$ in $cal_p$ of a service $p$ (as opposed to a component interface) 
implies $r$ to be provided by the component which calls $p$. 

Using a provided service $p$ in the $sub_r$ of a service $r$ but not in
the component interface, means that $p$ is accessible only 
during an interaction with $r$.

The behaviour ${\mathcal B}_s $ of a service $s$ is an 
\textit{extended labelled transition system} (eLTS)
defined by a 6-tuple 
 $\langle S, L, \delta, \Phi, S_0, S_F\rangle$ with 
$S$ the set of the states of $s$;
$L$ is the set of transition labels and  
$\delta$ is the transition relation ($\delta \in S \cross L \fun S$).
$S_0$ is  the initial state ($S_0\in S$),
$S_F$ is the finite set of final states ($S_F \subseteq S$),
$\Phi$ is a state annotation function ($\Phi \in S\fun sub_s$).
An eLTS is obtained when we allow nested states and transitions.
This provides a means to reduce the LTS size and a flexible description
with optional behaviours which take the form of sub-service names annotating some states.

\vspace{-0.2cm}
\paragraph{Transitions:}
The elements $((ss,label), ts)$ of $\delta$ have the concrete \kmelia syntax 
{\small \texttt{ss---label--->ts}}
where the labels are (possibly guarded) combinations
of actions: {\small\texttt{[guard] action*}}.
The actions may be \textit{elementary actions} or \textit{communication actions}.
An elementary action (an assignment for example) does not involve other services; it does not use a communication channel.
A communication action is either a \textit{service call/response} or a message \textit{communication}.
Therefore communications are matching pairs: \textit{send message(!)-receive message(?), call service(!!)-wait service start(??), emit service result(!!)-wait service result(??)}.
The \kmelia syntax of a communication action (inspired by the Hoare's CSP) is: 
{\texttt{channel(!|?|!!|??) message(param*)}}.

\vspace{-0.2cm}
\paragraph{Channels:}
A communication channel is established between the interacting services when assembling components. A channel defines a context for the communication actions.
At the moment one writes a behaviour, one does not know which components will communicate, but one has to know which channels will be used. A channel is usually named after the required service that represents the context. 
The placeholder keyword \texttt{CALLER} is a special channel that stands for the channel opened for a service call.
From the point of view of a provided service $p$, \texttt{CALLER} is the channel that is open when $p$ is called. From the point of view of the service that calls $p$, this channel is named after one of its required service, which is probably named $p$.
The placeholder keyword \texttt{SELF} is a special channel that stands for the channel opened for an internal service call.
In this case, the required service is also the provided service.

\section{Research Issues for Service-based Systems}
\label{section:researchIssue}
\vspace{-0.3cm}
 \subsection{The Basic Context}
The fundamental context of component or service-based system is that of
distributed, asynchronous models. In an asynchronous model a set of
processes cooperate to perform a task by exchanging messages over
communication support.
In some extent the context is that of distributed system with some
specificities due to services such as volatility, discovery  and dynamic
binding. Service-based may be studied as distributed ones \cite{LikeItorNotBirman2004}.

Correctness of component or service-based systems should be studied in
this context. The good news are that distributed systems have been well
researched, the bad news are that there are still challenging problems
that are now exacerbated by the popularisation of web-services and applications.

Numerous works have embraced formal verification of concurrent
(distributed) systems
\cite{DBLP:conf/issta/ChocklerFGGNR06,DBLP:conf/issta/WojcickiS06,WesselinkATVA2007,MuhlBook2006}.
The work in \cite{MorinAlloryAsynchronous2006} focus for example on
hardware but with asynchronous mode.
In \cite{Rychly2006,RychlySETP2008} the authors use the $\pi$-calculi to
reason about services and systems. These are works that can  support
rigorous analysis of service-oriented systems.

\vspace{-0.3cm}
\subsection{Service Construction}
First of all, as a service is a software entity that implements a given
functionality, its construction should follows the development
lifecycle of any software entity. It may be constructed by refinement
from an abstract and formal model. However the specificities of
distributed environment and particularly those of services may be taken
into consideration: 
An event-based modelling for example is well-adapted to asynchronous
systems. According to service features, the good abstraction level
that favours reuse, the appropriate interface for service description
and the specific characteristics of the current service should be the
parameters of the service construction.

The extension of service standard languages such  as WSDL and BPEL is
proposed.
WSDL may integrate property description features in order to embed the
properties inside the service interfaces. The extension may be good at  addressing
required working conditions description, functional and non-functional
properties description and, interaction properties to make them easier.

According to BPEL, the needed extensions may enable an easy checking of
composition after extracting/discovering  the interaction features.
Also an event-based interaction may facilitate loose composition.
There is a need of expressive formally defined languages an tools.
\vspace{-0.3cm}
\subsection{Service Composition}
As service-based systems are basically distributed systems, the
classical problems studied in the scope of distributed systems should be
considered again. The composition of services  to build larger applications may
cause deadlocks, access conflict to shared resources, race conditions,
starvation, etc.

Moreover, as services should be discovered before being used, planning the
composition of services is a challenging question.
To find the right services to be composed is a challenging issue
that raises again the question of \textit{specification matching}. Indeed one
has to search for a  service (which is unknown) with given criteria. The (partial) specification
of the needed service is one of these criteria. The desired properties
of the searched service are also candidates for the searching criteria. The
searching activities are then becoming very tedious. 
Today searching systems are based on a database of published services
which are known and explored by clients which retrieve the needed services.

Correctness of service-based systems requires the proof of global
properties by considering hypothesis on the future services that will be
found and bound.
That means, an analysis of a global system may be performed by
considering assumptions on what the required services will
be. Therefore, it remains to check that the concrete services satisfy
the assumptions.

As far as service composition is concerned, the interconnection between
services may be loose since service-based systems have dynamically
evolving architecture; this can be modelled and analysed using
event-based approaches \cite{coloss:CA_ICFEM06,coloss:CA_Isola2008}.

The extension of the languages like BPEL is a pragmatic way to
undertake the process of improving service composition. An extension
enabling one to check the ordering of events and also the temporal
properties of service interaction is of  real interest. 

\vspace{-0.3cm}
\subsection{Service Certification}

In order to ensure the quality of a published service which is devoted to be
used in building quality software, the certification of the service
quality is required. This may be achieved with respect to desired,
functional and non-functional properties. A service may be certified as
fulfilling given properties. That means the service implementation
satisfies its formal specification that states its properties.
This is the standard correctness issue required for software entities or
applications. However for the convenience of service context, we propose
that the services embed their properties so that they can be verified
by a third party entities, checked or considered for the integration of the
services in other environments. \\

\textit{\bf Property-carrying services} are therefore a solution towards the
insurance of quality.

A service may at least incorporate its
functional specification, the properties that it guaranties and the
properties that it assumes for a good functioning.

The client of a provided service which carries its properties  may be confident to the provided
service. Consequently an application that integrates several services
can coordinate and manage the properties required by ones and provided
by others. From this setting, reasoning on the entire system my be
undertaken. It is not straightforward to establish the
correctness in this context of the composition of asynchronous
(sub)systems. Works on Assume/Guarantee
\cite{xu98compositional,OwickiIsabelle99,MooijWesselink2003,giannakopouloAl2004} may be beneficially reused in
the component  or service-based systems.

A service should also be certified by a \textit{certifier} according to the
features it has declared. That means the certifier checks the properties
embedded in the service and in case of success certifies the
service. But the certifier should in turn be a sure one.
This matter may be managed by considering well-known provers,
efficient proof-checkers and specialised \textit{proof centers} with experts that perform
the proof on demand for the certification purpose.  Therefore only
the certificates coming from the adopted proof centers may be the confident
ones. We develop this idea of property-carrying services from the works
on Proof-Carrying Code originally elaborated by Necula \cite{NeculaPCC97}.

\vspace{-0.3cm}
\subsection{Service-based Application and Maintenance}
Service registries may change; services also may change due to
upgrades and evolution. Therefore clients of registries may be aware of the
modifications and should be able to discover incompatibilities if there
are any.
How one can be sure that a new version of a service is still correct
with its initial objectives. Here again, the certification approach may
be used. A service of a system may be replaced by another one without
breaking the initial objectives of the system.
The formal analysis with respect to hypothesis on services should
greatly help in doing this.
Service registries should be efficient and adaptable to any form of service
descriptions. They should offer the best retrieval features and
reasonable response time. 
Results from  database management systems should be exploited.

An important research direction is that of substituability between
services or components. A service could be replaced by another one
without breaking the chain of functionalities. The service binding
should be elaborated to permit such changes. Again, services equipped
with properties and certificates will favour this practice. 

\vspace{-0.3cm}
\subsection{Interoperability of Service-based Systems}
Abstraction is a cornerstone for interoperability. When various systems
are based on the same abstract view of the used entities and use predefined
rules to exchange and reason about entities of interest, ensuring
interoperability becomes easier. Consider for example the well-known Open
System Interconnection model of the ISO; it is the reference model that
ensures the interoperability between network protocols, operating
systems  and applications. Rules are defined at different levels of the
model, and standards are defined and respected by developers of devices,
protocols and  software. Few examples are the widely used Transport
Control Protocol (TCP), Simple Mail
Transfer Protocol (SMTP) or Hyper Text Transfer Protocol (HTTP).

The notions of services and service servers are already present
there. Each network service is clearly and uniformly identified by a
number that is used by all its clients. When a server is active, its
clients requests and accesses the desired services by using the
predefined usage manual of this service.

According to service-based systems, exploiting the know-how in the
networking, the experiment, methods and techniques of programming
distributed system are the main ingredients to ensure quality.
Practically, a multi-layered event-based design and development approach is recommended.
 
\vspace{-0.3cm}
\subsection{Reliability}
Correctness of services as software entities are not sufficient for
reliability. For a service-based system to be reliable, its environment
 should be taken into account: global correctness, i.e.
system level correctness should be considered. The availability of
network level services is then an important parameter of the
reliability. Permanent availability and good functioning of all the
components of the global system is required; obviously this cannot be
stated and systematically proved but assumptions may be made to ensure the acceptable
reliability according to the considered application. For example, failures of some components may be acceptable,
provided that the global system continues to offer a part of its
functionalities or to offer its functionalities with more delays.

How to check or detect failures of components/services of the system at hand? 
When a fixed number of interacting entities is considered, failure can
be detected by considering the monitoring of the entities. In case of
service-based systems which evolve dynamically as entities are found and
bound and unbound, the monitoring is not straightforward. Hence a
challenging issue.

Works on dynamically evolving systems, considering group membership and
distributed middle-ware systems may be beneficially used here. Ensemble \cite{BirmanIEEE2005,BirmanNetworkSOA05}
for example is a well-researched system that can be used for
service-based systems.

The QuickSilver\footnote{QuickSilver Scalable Multicast project website
at Cornell: www.cs.cornell.edu/projects/quicksilver/QSM/}
\cite{QuickSilver2006} system is also
a candidate to ensuring or studying reliability of web-service based systems.

We also advocate for a \textit{server of service servers} in order to facilitate access
and updates. Indeed for a client of services, instead of looking for
several service registries for example, it may always contact only one
server of the registries which in turn redirects the requests or accesses to
the right registries. It will be the role of the server of registries
to manage and maintain up-to-date the information concerning the registries.
This approach is more flexible and efficient than the former
where the clients should know and maintain their information
about several registries.  

 \vspace{-0.3cm}
\subsection{Security}
Several problems and solutions have been studied with respect to
Internet network applications. As service-based systems is based on
this support, the same problems impact these systems and the adopted
solutions are also available.
Service registries need specific care as they should be continuously
available. Technically a cluster of registries may fulfill the objectives.
For example, false service registries  may be avoided by considering
friend service registries. Inaccurate service publishing may be avoided
by considering well authenticated service publishers.

\vspace{-0.3cm}
\section{Concluding Remarks}
\label{section:conclu}
The dissemination of software services through information systems and
applications require now and quickly components and services of high
quality. Providing verifiable correct services is one solution but it
is only partial; more
generally, the reliability of component or  service-based systems is the
true challenge. It goes beyond the components and services correctness; it
deals with the environment of the used services, components and developed
applications. Layers of properties and correctness proofs are
required (services, service registries, components, client applications,
certifiers, etc). To follow the objectives of correctness and reliability,
it is worth considering the integration of correctness threats in the
design and the development of the building blocks which are services.
Current languages and techniques can be extended in this direction and
  they may also exploit existing results on formal analysis of
 concurrent distributed systems. We have indicated some research issues
 and work directions.
The magnitude of the task is important. But there are already a
considerable amount of foundation works that can be beneficially
exploited. Results in the field of distributed asynchronous systems
modelling and verification are of main interest.
Works on property-carrying services certification will provide
interesting assessments according to quality. However, efficient property
verification techniques, tools and verification centers will be a major
part of the effort. Language expressiveness and appropriate modelling
techniques, based on specific domains, will help to manage verification
complexity, by defining repeatable proof scenario that will accelerate
verification and diminish analysis and development cost. Last, efficient simulation environments will be of great help in tuning
service construction and service-based applications construction.    
\small
\bibliographystyle{plain}

\end{document}